\documentstyle[prl,aps,epsf]{revtex}
\input epsf
\begin{document}

\title{Complex Scaled Spectrum Completeness for Coupled Channels}

\author{B.G. Giraud}
\address{giraud@spht.saclay.cea.fr, Service de Physique 
Th\'eorique, DSM, CE Saclay, F-91191 Gif/Yvette, France}

\author{and}

\author{K. Kat\=o and A. Ohnishi}
\address{kato@nucl.sci.hokudai.ac.jp, ohnishi@nucl.sci.hokudai.ac.jp, 
Division of Physics, \\
Graduate School of Science, Hokkaido University, Sapporo 060-0810, Japan}

\date{\today}
\maketitle

\begin{abstract}

The Complex Scaling Method (CSM) provides scattering wave functions which
regularize resonances and suggest a  resolution of the identity in terms
of such resonances, completed by the bound states and a smoothed continuum.
But, in the case of inelastic scattering with many channels, the existence 
of such a resolution under complex scaling is still debated. Taking advantage 
of results obtained earlier for the two channel case, this paper proposes 
a representation in which the convergence of a resolution of the identity can 
be more easily tested. The representation is valid for any finite number of 
coupled channels for inelastic scattering without rearrangement.

\end{abstract}

\section{Introduction, Notations}

\medskip
As is well known, the CSM converts the description of resonances by 
non-integrable Gamow states into one by square integrable states while 
leaving the discrete spectrum unchanged \cite{ABC}. Cuts describing the 
continuum are rotated, however, but this may be advantageous, since they are 
thus disentangled when their thresholds differ from one another. (We are not 
interested, in this paper, in the case of channels with identical thresholds.)
It is then expected that the continuum corresponding to such rotated cuts 
makes a much smoother contribution to the calculation of collision amplitudes, 
level densities, strength functions and sum rules \cite{Kato1} \cite{Kato2}, 
since narrow resonant processes have been assumed to be peeled out explicitly 
by the CSM. The CSM Hamiltonian, unfortunately, is not hermitian any more, 
and it is not obvious that a resolution of the identity in terms of its bound 
states, resonances and presumably damped continuum is possible. For the one 
channel case, convincing arguments have been advanced a long time ago 
\cite{Berg1} to prove that this resolution exists. More recently \cite{us}, a 
detailed investigation of the case of two channels, coupled by straightforward 
potentials, generated a contour integration of the usual Green's function 
which provided the identity resolution. The task was made reasonably easy by 
the small complication of the Riemann surface in that case. The purpose of the 
present paper is to capitalize on the methods used for that two channel case 
and attempt a generalization to any finite number of channels, despite the
more complicated nature of the relevant Riemann surface. We shall assume, 
naturally, that there already exists, derived from single poles and usual 
cuts, a resolution of the identity for the initial Hamiltonian, before its 
modification by complex scaling. Our problematics would be meaningless 
otherwise.

\medskip
Several earlier studies, in particular by \cite{KFFGR} \cite{ML}, have been 
concerned with a description of resonances with square integrable states, 
without complex scaling. They did not restrict to the consideration of just
simple poles of the $S$-matrix and investigated how one might, as rigorously 
as possible, define initial wave packets for the description of decaying 
states; the non purely exponential nature of their decays received a detailed 
attention, via the analysis of their time dependent evolutions. The present 
paper, however, will be content with a Gamow definition of resonances, by 
means of simple poles; our aim is just to generate a resolution of the 
identity, with time independent states extending to asymptotic regions. For 
earlier searches of a complete basis of states, including resonances, but 
within a compact interaction volume, we may refer to the review by \cite{BRT} 
of $R$-matrix methods and their extensions. See also \cite{R} and in particular
the comparison of ``class B'' and ``class D'' theories.

\medskip
In this paper, we shall again assume that all potentials $V_{in}(r)$ driving
the channels and their couplings are local and so short ranged, Gaussian-like 
for instance, that the $2N$ Jost solutions of the $N$ coupled equation system,
\begin{equation}
-\psi_{ij}''(k_j,r)+\sum_{n=1}^N  
\left[ e^{2i\theta} V_{in}(e^{i\theta}r)+
\left(\frac{\ell_i(\ell_i+1)}{r^2}-k_i^2\right) \delta_{in} \right] 
\psi_{nj}(k_j,r)=0,\ \ i,j=1,...,N,
\label{coupl}
\end{equation}
exist and are analytical in the whole complex domain of all the momenta $k_j.$
The radius $r$ runs from $0$ to $+\infty,$ obviously, and the number $N$ of 
channels is taken as finite. As an additional technicality we also assume, 
naturally, that the products $V_{in}\, \psi_{nj}$ do not diverge for 
$r \rightarrow 0$ when singular solutions of Eqs.(\ref{coupl}) are considered.

\medskip
We select the threshold of the lowest channel as the origin of the complex 
energy plane, hence $E \equiv k_1^2.$ The other channels with their 
physical thresholds $E_j^*,$ which are real and positive numbers, now 
define channel momenta according to, $E_j=k_j^2=E-e^{2i\theta}E_j^*.$ Notice 
that, given a real number $E_j^*$ defining a physical threshold, the usual 
complex scaling where $p^2$ becomes $e^{-2i\theta} p^2$ and $r$ becomes 
$e^{i\theta} r$ does not change $E_j^*$ and rotates the corresponding cut by 
angle $-2\theta.$ But here, we have a slightly different representation, 
because the Hamiltonian has been multiplied by $e^{2i\theta}.$ Hence kinetic 
operators in our Hamiltonian $H,$ see Eqs.(\ref{coupl}), are just $-d^2/dr^2,$ 
every cut rotates back into being ``horizontal''and starts from 
$e^{2i\theta} E^*.$ For time dependent studies. it will make sense to scale 
time, conjugate of energy, by a factor $e^{-2i\theta}.$ This will prevent 
those resonant wave packets, the energies of which have a positive imaginary 
part as eigenvalues of $H,$ from exploding when $t \rightarrow + \infty.$

\medskip
Also in this paper no rearrangement is allowed, channels are defined by 
just internal excitations of the projectile and/or the target, hence all 
reduced masses are equal. Finally we exclude from this paper the consideration
of abnormal thresholds; we shall only discuss the case of ``square root 
thresholds''. This is generic enough.

\medskip
It is understood here and from now on that a first subscript, such as $i$ or 
$n,$ denotes the component of each wave $\psi$ in channel $i$ or $n,$ then 
that any superscript, $\pm,$ or second subscript, $j,$ denotes the boundary 
condition which defines $\psi.$ For a Jost solution $f^{\pm}_{.j},$ the 
boundary condition that we choose is ``asymptotic flux 
$e^{\pm i(k_jr-\ell_j \pi/2)}$ in channel $j$ and no asymptotic flux in the 
other channels''. It is well known that for $r \rightarrow 0,$ the components 
of such Jost solutions are proportional to $(k_ir)^{-\ell_i}(2\ell_i-1)!!.$ 
For a regular solution $\varphi_{.j},$ the boundary condition that we choose 
sits at $r=0$ and reads, 
``$\lim_{r \rightarrow 0}\, (k_i r)^{-\ell_i-1}\, \varphi_{ij}(r) = 0\ \, 
\forall i \ne j,$ while, for $i=j,$ then 
$\lim_{r \rightarrow 0}\, (k_j r)^{-\ell_j-1}\, \varphi_{jj}(r) =
1/(2\ell_j+1)!!.$ 

\medskip
Following Newton \cite{Newt}, it is convenient, given $E$ and $r,$ to set the 
column vectors $\varphi_{.j}$ into a matrix ${\bf \Phi}(E,r)$ of regular 
solutions and the Jost solutions $f^+_{.j}$ (resp. $f^-_{.j}$) into a similar 
matrix ${\bf f}^+(E,r)$ (resp. ${\bf f}^-$). 
It is also convenient to notice that ${\bf \Phi}$, viewed as a function 
of the $k_j$'s as if these were independent momenta, is even under any 
reversal of a $k_j$ into $-k_j.$ Such is not the case for ${\bf f}^+;$ 
analytic continuations in either energy or momenta planes can introduce 
one (or several) $f^-_{.j}$'s into ${\bf f}^+.$
 
\medskip
For our oncoming argument we must use the Wronskian matrix with matrix 
elements the Wronskians ${\cal W}\left(f^+_{.m},\varphi_{.n}\right)$ of the 
Jost solutions $f^+_{.m}$ with the regular ones $\varphi_{.n}.$  This, for 
$s$ waves, is the transposed of ${\bf f}^+$ at $r=0,$
\begin{equation}
{\bf W}(E)=\tilde{{\bf f}}^+(E,0),
\label{witch}
\end{equation}
and for other angular momenta is only a slight modification of 
$\tilde{{\bf f}}^+(E,0).$ (Rather than just $\tilde{{\bf f}}^+(E,0)$ one 
must use limits of products $(k_ir)^{\ell_i}  f_{ij}^+/(2\ell_i-1)!!$ 
at $r=0,$ explicitly, but we will disregard this technicality.)
The Green's function ${\bf G}$ is then found as,
\begin{equation}
{\bf G}(E,r,r')={\bf \Phi}(E,r)\, [{\bf W}(E)]^{-1}\, \tilde {\bf f}^+(E,r')
\ \ {\rm if}\ r < r',\ \ \ 
{\bf G}(E,r,r')={\bf f}^+(E,r)\, [\tilde {\bf  W}(E)]^{-1}\, \tilde {\bf \Phi}
(E,r')\ \ {\rm if}\  r > r'.
\label{Green}
\end{equation}
Here each tilde $\tilde{ }$ means transposition; we refer to \cite{Newt}
or to Appendix A of \cite{us} for the derivation of such formulae for 
${\bf G}.$ Despite different formulae whether $r > r'$ or $r < r',$ and the 
lack of hermiticity, ${\bf G}$ is symmetric, namely 
${\bf G}(r,r')={\bf G}(r',r).$

\medskip
It will be noticed that the CSM, as we describe it by the system of 
Eqs.(\ref{coupl}), locates thresholds on a segment of the complex $E$ plane 
with slope $2\theta,$ extending from $E=0$ to $e^{2i\theta}E_N^*,$ and 
that the channel cuts are rotated back into being ``horizontal''. Conversely, 
bound states lie on a negative semiaxis rotated by $2\theta$ and resonances 
are rotated by $2\theta$ as well. This slight change of representation changes 
nothing to the physics, obviously. For trivial technical reasons \cite{us}, 
we normalize energy units so that $E_N^*=4.$ Also we shall  use a short 
notation, $k \equiv k_1$ and $K \equiv k_N.$ We show in Figure 1 the cut 
energy plane in an illustrative, four channel situation when $\theta=\pi/6,$ 
$E_2^*=1.5$ and $E_3^*=3.5.$

\medskip
Equipped with this slightly unwieldy formalism, we can now investigate whether
there exists a representation, and an integration contour, such that the 
traditional integral, ${\cal I}=\int dE\, {\bf G}(E,r,r'),$ calculated in two 
different ways, generates a resolution of the identity. This question of a 
representation and a contour is the subject of Section II, the main part of 
our argument. Additional considerations on the two ways of calculating this 
integral make the subject of Section III. A discussion and conclusion are 
proposed in Section IV. 

\begin{figure}[htb] \centering
\mbox{  \epsfysize=110mm
         \epsffile{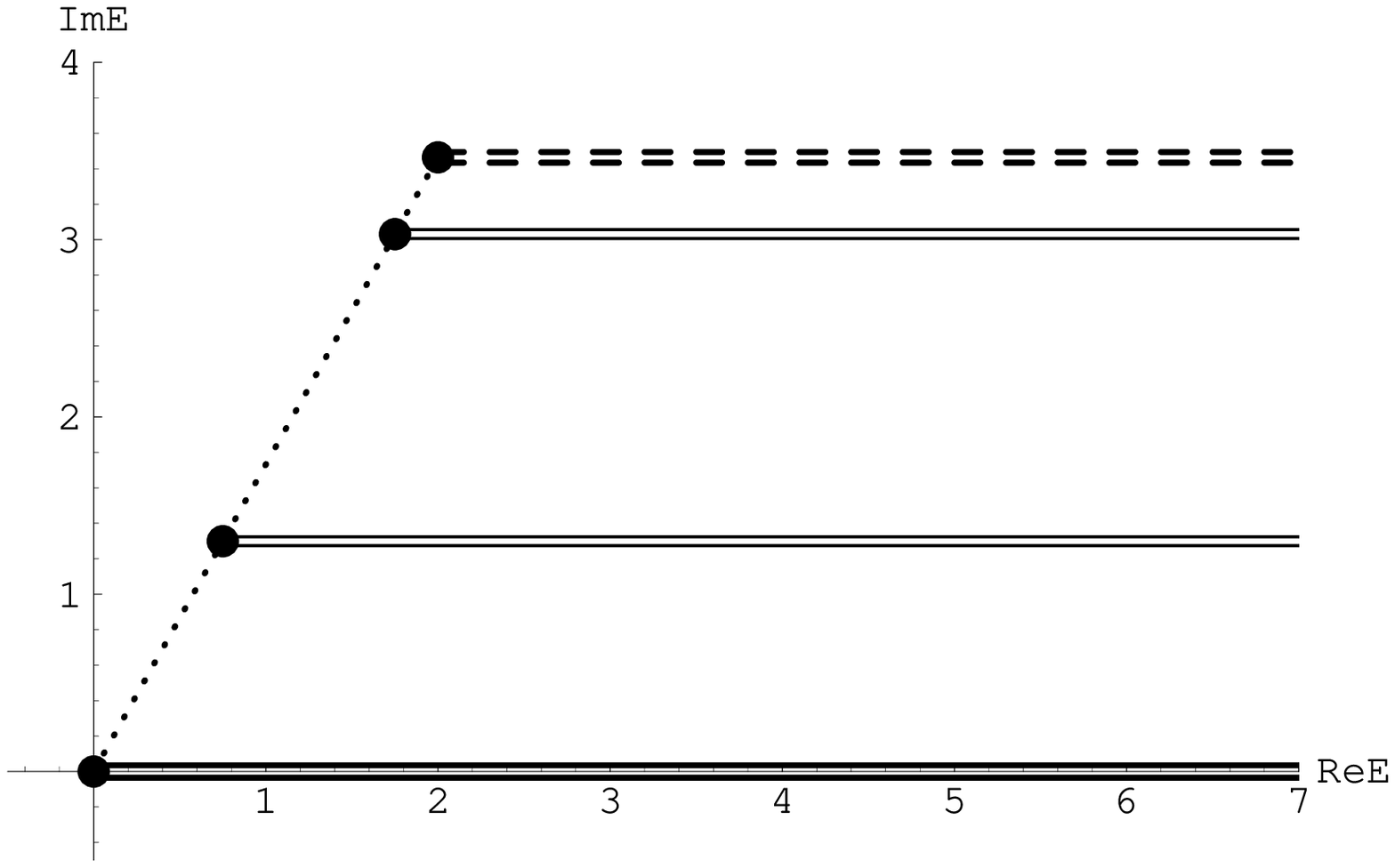}
     }
\caption{$E$-plane. Physical cuts for a four channel case when $\theta=\pi/6,$ 
$E^*_2=1.5,$ $E^*_3=3.5$ and $E^*_4=4.$ Lowest channel, heavy full lines, 
highest channel, heavy dashed lines, intermediate channels, lighter full 
lines. The dotted segment with slope $\pi/3$ is the locus of thresholds 
(big dots) in this representation.} 
\end{figure}

\section{Representations and contours}

\subsection{Energy plane}

\medskip
From Fig. 1 it is intuitive that one could start, for instance, from 
$+ \infty$ along the lower rim of the lowest channel cut, return to the origin,
$E=0,$ proceed to $+ \infty$ again on the upper rim, then join there the lower 
rim of the second cut, return to the threshold of this second cut, go to 
$e^{2i\theta}E^*_2+ \infty$ along the upper rim, join the third cut lower rim 
at infinity, etc., until arriving at $e^{2i\theta}E^*_N + \infty$ along the 
upper rim of the highest channel. Then the contour would be closing at infinity
by means of an almost complete circle, counterclockwise, terminating at the 
starting point, namely at $+ \infty$ on the lower rim of the lowest channel.

\medskip
Along such a contour, it would be necessary to investigate the behaviors of the
ingredients ${\bf f}^+,$ ${\bf W}$ and ${\bf \Phi}$ of ${\bf G}.$ Furthermore, 
information is needed about the singularities of ${\bf G}$ inside the contour;
indeed, residues of simple poles are essential for a calculation of 
$\int dE\, {\bf G}(E)$ by Cauchy's theorem; one also needs reasons why no 
singularities higher than simple poles occur.

\medskip
The representation discussed in the next subsection makes easier the needed 
investigation, for it opens two of the cuts and limits the discussion to
situations where all momenta have semipositive imaginary parts, 
$\Im k_j \ge 0.$

\subsection{Pseudomomentum plane}

\medskip
A generalization from \cite{us}, where there were two channels only, the 
present ``$P$ representation'' consists in joining the upper rim of the 
lowest cut and the lower rim of the highest cut, and in opening both cuts, 
by  {\it rational} formulae,
\begin{equation}
k=P+Q^2/P,\ \ K=P-Q^2/P,
\end{equation}
where $Q=e^{i\theta}$ makes a short notation for our scaling of energies such
that $E^*_N=4$ and $k^2-K^2=4Q^2.$ Trivially, $P$ is the average $(k+K)/2$ of 
$k$ and $K.$ The point is, despite an obvious failure to open additional cuts,
$P$ also give the ``dominant'' part of any other momentum when 
$\Im P \rightarrow + \infty.$
Indeed, when $|P|$ is large, say $|P| >> 2,$ then an asymptotic value can be 
defined for $k_j,\, j \ne 1,\, j \ne N,$ according to the rule, 
\begin{equation}
k_j \equiv (k^2-Q^2 E^*_j)^{\frac{1}{2}} = 
(P^2+2Q^2-Q^2 E^*_j+Q^4/P^2)^{\frac{1}{2}} = P + Q^2(1-E^*_j/2)/P + 
{\cal O}(P^{-2}).
\end{equation}
Thus the semicircle at infinity in the upper $P$ plane corresponds to 
$\Im k_j >0,\, \forall j.$ This is of critical value for the zoology
of our Jost functions and it is expected that this semicircle properly
closes the integration contour under design.

\medskip
Set now $P=x+iy$ and short notations $c=\cos2\theta$ and $s=\sin2\theta.$
A trivial calculation separates the real and imaginary parts
of the (complex) energies driving each channel,
\begin{equation}
(x^2+y^2)^2\, \Re(k_j^2)=
[(x^2+y^2+s)(x+y)+(x-y)c] \, [(x^2+y^2-s)(x-y)+(x+y)c] 
- E_j^* (x^2+y^2)^2 c,
\label{rea}
\end{equation}
and
\begin{equation}
(x^2+y^2)^2\, \Im(k_j^2)=2[(x^2+y^2) x + x c + y s]\, [(x^2+y^2) y + x s - y c]
- E_j^* (x^2+y^2)^2 s.
\label{ima}
\end{equation}
and it is trivial to recover the images, in this new representation, of the 
cuts displayed in Fig. 1. Polar coordinates, with $P=p e^{i\eta},$
can be also be used to decribe the $j$-th cut from Eq.(\ref{ima}) by, 
\begin{equation}
p^2 \sin2\eta + \frac{\sin(4\theta-2\eta)}{p^2}=(E^*_j-2)\, \sin2\theta.
\label{polar}
\end{equation}

\begin{figure}[htb] \centering
\mbox{  \epsfysize=110mm
         \epsffile{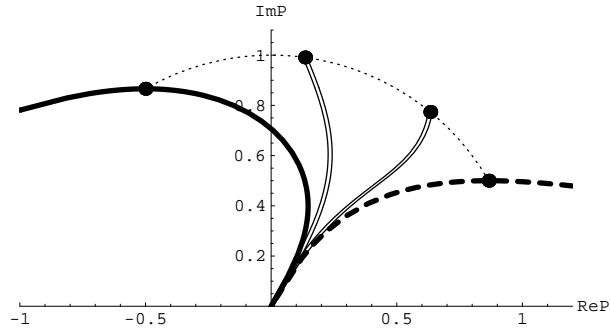}
     }
\caption{$P$ plane. Cuts for the same four channel case, $\theta=\pi/6,$ 
$E^*_2=1.5,$ $E^*_3=3.5$ and $E^*_4=4.$ Opened cut for lowest channel, 
heavy full line. Opened cut for highest channel, heavy dashed line. 
Intermediate channel cuts, not open, lighter full lines. The dotted segment 
is the locus of thresholds (big dots) in this $P$ representation.} 
\end{figure}

Results are shown in Figure 2 for the same special case as Fig. 1.
As in \cite{us}, the lowest channel is represented by the heavy,
shoulder shaped line, that starts from $-\infty$ on the real P axis, 
bends up, then backs into the origin $P=0,$ where it terminates with a slope 
$2 \theta.$ Along the curve, $k$ is real and runs from $-\infty$ to $+\infty,$
covering both rims of the initial cut. The threshold $k=0$ is represented by
$P=iQ=e^{i(\theta+\frac{1}{2}\pi)}.$ Partner points where 
$k \leftrightarrow -k$ obtain under the symmetric transformation 
$P \leftrightarrow -Q^2/P.$ In the same way, for the highest channel, $K$ runs
with real values along the heavy dashed line, from $-\infty$ at $P=0$ to 
$+\infty$ at the end of the positive $\Re P$ semiaxis, via $K=0$ for $P=Q.$ 
The transform, $P \leftrightarrow Q^2/P,$ makes partners with opposite values 
of $K.$

\medskip
The other cuts remain cuts. Their thresholds lie on the image, shown as a 
dotted line again, of the segment already pointed out at the stage of Fig. 1.
Because both $\Re (k_j^2)$ and $\Im (k_j^2)$ vanish for such points, it is 
easy to eliminate $E_j^*$ between the right hand sides of 
Eqs.(\ref{rea},\ref{ima}) and obtain the condition for such a locus,
\begin{equation}
x^2+y^2=1,
\end{equation}
a very simple result indeed. With $|P|=1,$ the positions of the thresholds are 
easy to obtain. The special cases $j=1$ and $j=N$ give the argument
$\eta \equiv ArgP$ as $\eta=\theta+\pi/2$ and $\eta=\theta,$ respectively. 
This was already known from \cite{us}. The function 
$\sin2\eta+\sin(4\theta-2\eta),$ see Eq.(\ref{polar}), decreases monotonically 
when $\eta$ increases from $\theta$ to $\theta+\pi/2,$ hence a unique solution 
for each $E_j^*,$ and an obvious symmetry about $\theta+\pi/4$ corresponding 
to the symmetry about $E_j^*=2.$ Then each intermediate cut generates, from 
Eq.(\ref{ima}), an image which joins its threshold to the origin $P=0,$ while 
$k_j,$ a real number along this image, runs from $0$ to $\pm \infty,$ 
according to the rim. The image lies between the heavy full and dashed lines, 
and, being pinched between them at $P=0,$ also reaches the origin with slope 
$2 \theta.$ While the pinching makes numerics slightly difficult, it is easy 
to verify analytically from Eqs.(\ref{rea},\ref{ima}) that {\it 
infinitesimally away from both rims of such an intermediate cut, but inside 
the wedge created by the heavy line curves, $\Im k_j$ remains positive}.

\begin{figure}[htb] \centering
\mbox{  \epsfysize=110mm
         \epsffile{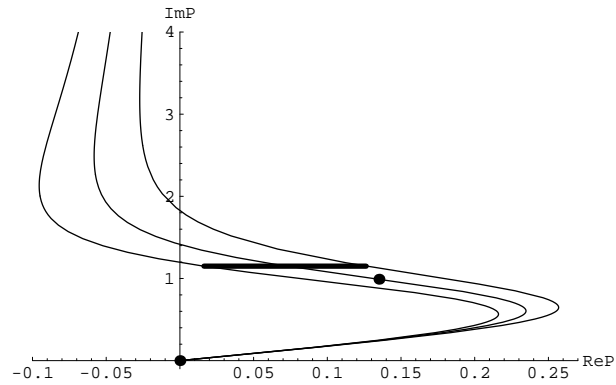}
     }
\caption{$P$ plane. Again  $\theta=\pi/6.$ Cut for the channel defined by
$E^*_2=1.5.$ The center line, between dots, is the cut. Cut then continued
for negative energies in the channel. Additional lines, lower rim (leftmost 
curve) and upper rim (rightmost curve), respectively. Both rims extended 
below threshold. Heavy line bar, connection between extended rims.}
\end{figure}

To illustrate our full control of the various $\Im k_j$'s provided by this $P$
representation, whether inside the wedge or near the positive infinity 
semicircle, we show in Figure 3 the cut corresponding to $E_2^*,$ and its 
continuation beyond threshold.  By ``beyond'', we mean still canceling 
$\Im k_2^2,$ while $\Re k_2^2$ becomes more and more negative. This allows 
reaching the ``semicircle''. Simultaneously, we generate rims of the cut, and 
beyond again below threshold. To generate rims, we use Eq.(\ref{ima}), or as 
well Eq.(\ref{polar}), with $E_2^*$ replaced by $E_2^*-0.2$ and $E_2^*+0.2$ 
for the lower and upper rim, respectively. (The choice $\pm 0.2$ was made for 
graphical convenience, but we tested much smaller intervals, naturally.) The 
dots represent $P=0,$ where the channel energy is infinite, and the threshold,
where it vanishes by definition. Like the cut, the rims are pinched by the 
wedge. 

Then we show in Figure 4 the trajectory of $k_3$ when $P$ follows this
cut from $P=0,$ to the threshold and beyond. Notice that, $E_2$ being real 
along the line, then the imaginary part of $E_3=E_2+e^{2i\theta}(E_2^*-E_3^*)$ 
is obviously negative. This does not prevent a choice of $k_3$ with 
$\Im k_3>0,$  generating the leftmost trajectory in Fig. 4.  
Simultaneously, we show the trajectories of $k_2$ from both rims of the same 
cut. The left hand side (when seen in Fig. 3) rim induces 
$\Re k_2 \rightarrow -\infty$ when $P \rightarrow 0,$ with an 
infinitesimally positive $\Im k_2.$ Conversely the right hand side rim 
induces $\Re k_2 \rightarrow +\infty$ when $P \rightarrow 0,$ with still 
an infinitesimally positive $\Im k_2.$ When we go from either rim towards the 
upper semicircle at infinity, this induces $\Im k_2 \rightarrow + \infty,$
as expected. The rims can be connected by any small path, see the bar
above the threshold in Fig. 3, and the values of $k_2$ along the rims can be 
smoothly matched, see the curved bar in Fig. 4, the trajectory of 
$k_2$ when $P$ follows the bar in Fig. 3. Generalizations to every $k_j$ in 
every part of the wedge are trivial.

\begin{figure}[htb] \centering
\mbox{  \epsfysize=110mm
         \epsffile{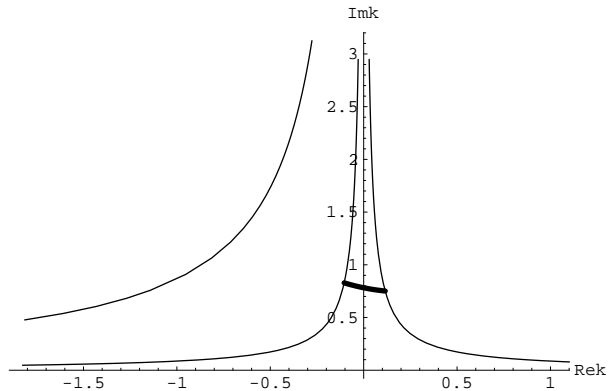}
     }
\caption{$k_2,k_3$ planes. Still  $\theta=\pi/6,$ $E_2^*=1.5$ and $E_3^*=3.5.$ 
Leftmost curve, trajectory of $k_3$ when $P$ follows the central line of 
Fig. 3. Intermediate curve, trajectory of $k_2$ for extended lower 
rim, see leftmost curve in Fig. 3. Rightmost curve, trajectory of $k_2$ 
induced by extended upper rim, see rightmost curve in Fig. 3. Heavy line 
curved bar, connection trajectory for $k_2$ when $P$ turns around the 
threshold, below it.}
\end{figure}

\subsection{Contour}

\medskip
To synthetize this Section, the $P$ representation defines a physical sheet 
similar to the physical sheet of the energy plane. The region of interest is
that region above the two curves which open the cuts for the lowest and the 
highest channels, while cuts remain for the intermediate channels. All momenta 
inside the wedge, and all the way to the upper semicircle at infinity, can 
be defined with positive imaginary parts. A contour can be found, following
all cuts and closing at infinity in the upper plane.

\medskip
The intuition which was present in the $E$ representation can be 
substantiated in the $P$ plane. Start from $- \infty$ on the real axis, follow
the ``opener curve'' which corresponds to the lowest channel, all the way
to $P=0.$ From there, follow the lower rim of the cut corresponding to the 
second channel, back to its threshold, then turn around the threshold to 
follow its upper rim, down to $P=0.$ In turn, follow the lower rim of each 
intermediate channel, then its upper rim. After bouncing $N-1$ times at 
$P=0,$ follow the ``opener curve'' corresponding to the upper channel, until
$P \rightarrow + \infty$ on the real axis. Then close the contour by means of
the upper semicircle at infinity. In the next Section, we shall investigate 
what happens to the integral, ${\cal I}=\int dE\, {\bf G}(E,r,r'),$ when 
considered along this contour in the $P$ plane.

\section{Three contributions to the Green's function integral}

\subsection{Upper semicircle}

\medskip
At infinity in this upper $P$ plane, the integration weight, 
$dE=2\left(P-Q^4/P^3\right)\,dP,$ boils down to $2P\,dP.$  All the $N$ 
distinct Jost solutions boil down to 
$\exp\left[i(Pr-\frac{1}{2}\ell_j\pi)\right]$ in their 
respective ``flux channel $j$'', while vanishing in the other channels. At 
the same time, the $N$ distinct regular solutions similarly boil down to 
$\sin\left(Pr-\frac{1}{2}\ell_j\pi\right)/P$ in their respective flux channel 
and vanish in the other channels. The Wronskian matrix boils down to the 
$N$-dimensional unit matrix.

\medskip
Assume now $r > r',$ for instance, and thus consider the second of 
Eqs.(\ref{Green}). The product 
${\bf f}^+\, [\tilde {\bf W}]^{-1}\, \tilde {\bf \Phi}$ boils down to a
diagonal matrix. Its $j$-th diagonal element reads,
\begin{equation}
\int_{sc}2\ dP\ e^{i(Pr-\ell_j\pi/2)}\,
\sin\left(Pr'-\ell_j \frac{\pi}{2}\right),
\end{equation}
and can be easily calculated by reducing the semicircle back to the real $P$
axis. The result does not depend on $j,$
\begin{equation}
-i \int_{\infty}^{-\infty}\, dP\, e^{i(Pr-\frac{1}{2}\ell_j\pi)}\,
\left[ \exp\left(iPr'-i \ell_j \frac{\pi}{2}\right) -
       \exp\left(i \ell_j \frac{\pi}{2}-iPr'\right) \right] =
2 i \pi [\delta(r+r') - \delta(r-r')].
\end{equation}
It is trivial to verify that the same result is obtained if $r<r'.$ 
Furthermore the term $\delta(r+r')$ cancels out in the space of regular radial 
waves. Hence the contribution ${\cal I}_{sc}$ of the semicircle makes nothing 
but the multichannel identity, multiplied by  $(-2i\pi).$ Notice that, 
differing from \cite{Kato1}, this identity is not multiplied by a factor 
depending on $\theta,$ since for us the ends of the semicircle, $-\infty$ and 
$+\infty,$ both lie on the real $P$ axis.

\subsection{Continuum}

It makes no difference here whether we consider the contribution of one of the
``opener line'' or that of one of the intermediate cuts. For in both cases
we group partner terms. Such partners either come from a transform 
$P \leftrightarrow \pm Q^2/P$ or from opposite rims of the intermediate cut
under consideration. 
What is important to notice is that momenta retain their finite and positive 
imaginary parts and do not change when we compare two partner points, except 
that momentum specific to the opener line or the cut. For that momentum, which 
is real, ``partnership'' means $k_j \leftrightarrow -k_j,$ with still an 
infinitesimal positive imaginary part. Keeping in mind 
that ${\bf \Phi}$ is even under such a momentum flip, the contribution of
such a continuum thus reads, if $r > r'$ for instance,
\begin{equation}
{\cal I}_j=\int_0^{\infty} 2k_j\, dk_j\,  {\bf D}_j(E,r)\, 
\tilde {\bf \Phi}(E,r'),
\label{cut}
\end{equation}
where ${\bf D}_j(E,r)$ represents the following difference between partners,
\begin{equation}
{\bf D}_j(E,r)={\bf f}^+(E,   r)\, [\tilde {\bf W}(   E)]^{-1} -
               {\bf f}^+(-k_j,r)\, [\tilde {\bf W}(-k_j)]^{-1},
\label{discont}
\end{equation}
a discontinuity across the cut. The notation used here takes advantage of
the fact that $dE=2k_j dk_j,$ and that $k_j$ is a convenient label along
the line or the cut. 
The first term, ${\bf f}^+(E,r)\, [\tilde {\bf W}(E)]^{-1},$ 
in the right hand side of Eq.(\ref{discont}) clearly comes from the upper rim. 
The notation that we use for the second term, 
${\bf f}^+(-k_j,r)\, [\tilde {\bf W}(-k_j)]^{-1},$ indicates that, because of 
analytic continuation in the physical sheet around the threshold, one Jost 
solution $f^-_{.j}$ now makes the $j$-th column of ${\bf f}$ and that of 
$\tilde {\bf W}.$ {\it All other columns are unchanged, and this strong 
similarity reduces the difference ${\bf D}_j$ to be a rank one dyadic.}
An elementary proof of this dyadic result was given in Appendix C of \cite{us}.
Nothing changes in the argument if $r < r'.$

\medskip
As a consequence of the dyadic nature of ${\bf D}_j,$ and of the symmetry  
${\bf G}(E,r,r')={\bf G}(E,r',r),$ hence of the same symmetry for 
discontinuities across cuts, there exists as a column vector a solution 
$\phi_{.j}$ of Eqs.(\ref{coupl}) that is able to represent symmetrically
both ${\bf D}_j(E, r)\, \tilde {\bf \Phi}(E,r')$
and
${\bf \Phi}(E,r)\, \tilde {\bf D}_j(E,r')$
in a self dual way as an outer product,
\begin{equation}
{\cal I}_j=\int_0^{\infty} 2k_j\, dk_j\,  \frac{\phi_{.j}(E,r)\, 
{\tilde \phi}_{.j}(E,r')}{{\cal D}(E)}.
\label{continuum}
\end{equation}
This solution belongs to the set of regular solutions, naturally, because of
the regularity of ${\bf G}$ at both $r=0,$ and $r'=0,$ illustrated by the 
presence of ${\bf \Phi}$ in Eqs.(\ref{Green}). 
The exact natures of this $\phi_{.j}$ and of the ``normalizing'' 
denominator ${\cal D}$ are discussed in the Appendix.

\medskip
At this stage, the full integral along the full contour thus gives the 
sum of the multichannel identity and ``pseudoprojectors on the continuum'', 
one pseudoprojector for each channel,
\begin{equation}
\frac{i}{2\pi}\int dE\, {\bf G}(E,r,r')=
\left[\matrix{\delta(r-r') & 0 & ... & 0 \cr
               0 & \delta(r-r')& ... & 0 \cr
               . &     .       &  .  & . \cr
               0 &     0 & ... & \delta(r-r')}\right] + 
\frac{i}{\pi}\sum_{j=1}^N
\int_0^{\infty} k_j\, dk_j\, \frac {\phi_{.j}(E,r)\, {\tilde \phi}_{.j}(E,r')}
{{\cal D}(E)}.
\label{resola}
\end{equation}
The next subsection shows what happens if the same integral is evaluated
by means of the Cauchy theorem.

\subsection{Residues at poles}

\medskip
We assumed that, before complex scaling, namely for $\theta=0,$ there existed 
an identity resolution in terms of unscaled bound states and unscaled 
scattering states. In other words we assumed that the corresponding, unscaled 
${\bf G}(E)$ shows only isolated, simple poles, besides the physical cuts. 
Such poles can be on the real $E$ axis of the physical sheet, describing 
bound states, or away from this axis, then describing resonances or 
antiresonances. The point is, now, that the CSM cannot change the nature of
such poles \cite{ABC}. Within our description by Eqs.(\ref{coupl}), the CSM
just rotates such poles by $2\theta$ in the energy representation, 
along circular arcs, concentric around $E=0.$ In the $P$ representation, 
the images of such arcs are also concentric arcs, with angular extension 
$\theta$ only. This is trivially seen from the equation which, for each 
initial position $\varepsilon$ of a pole, defines those values of $P$ which 
represent $e^{2i\theta}\varepsilon,$
\begin{equation}
\left(P+e^{2i\theta}/P\right)^2=e^{2i\theta}\, \varepsilon.
\end{equation}
Indeed, $\theta$ disappears from this equation if one sets $P=e^{i\theta}P_0,$ 
where $P_0$ solves for the initial position $\varepsilon.$ It can be concluded
that only simple poles will be found when a finite $\theta$ is used
for our CSM. Notice, incidentally, that for $\varepsilon$ real and negative
(bound states), the $P$ representation will align poles along the axis with
polar angle $\theta+\pi/2,$ further than the circle with radius $1$ that we
found as the locus of thresholds. There will be no such alignment for 
resonances.

\medskip
For the calculation of ${\cal I}$ by Cauchy's theorem, poles are not due to 
either ${\bf f}$ or ${\bf \Phi},$ since these, as functions of $E$ or $P,$ are 
regular. Only the divergence of ${\bf W}^{-1}$ can create poles. The 
situations of interest are those when the roots of the determinant, 
$\det {\bf W},$ are located inside the integration contour. We know that 
such is the case for the bound states. Depending upon $\theta,$ some 
resonances may also rotate into the domain. It is  already known that only 
simple, isolated poles occur. The only question to solve is, what is the 
residue of ${\bf G}$ at such a pole.

\medskip
Residues of ${\bf G}$ at its poles will now be obtained from derivatives 
$d/dE.$ That is equivalent to a calculation in the $P$ representation, anyhow, 
and slighly easier. We shall use short notations in which the dependence of 
${\bf \Phi},$ ${\bf f}^+,$ ${\bf W},$ upon $r,$ and/or $r'$ and/or $E$ will be 
most often understood. However, at those energies $E_{\nu}$ 
where a pole occurs, we use an explicit subscript $\nu$ to specify that such 
quantities ${\bf \Phi},$ ... ,  ${\bf W}$ are evaluated at $E_{\nu}.$

\medskip
Poles occur because of ${\bf W}^{-1}.$ Hence, we must only find the residue,
\begin{equation}
{\cal R}_{\nu}=\lim_{E \rightarrow E_{\nu}}\ (E-E_{\nu})\ {\bf W}^{-1}(E)\, ,
\label{residue1}
\end{equation}
and form the matrix product, 
${\bf \Phi}_{\nu}\,{\cal R}_{\nu}\,\tilde {\bf f}^+_\nu$ and its
transpose ${\bf f}^+_\nu\,\tilde {\cal R}_{\nu}\,\tilde {\bf \Phi}_{\nu}.$ 

\medskip
At a (simple!) root $E_{\nu}$ of $\det{\bf W}(E),$ there is necessarily one, 
and just one, null right eigenvector $\Lambda_{\nu}$ of ${\bf W}.$ Similarly 
there is one, and just one, null left eigenvector $\Lambda'_{\nu}.$ We write 
them as columns and normalize them by the condition,
\begin{equation}
\tilde \Lambda'_{\nu}\, \Lambda_{\nu}=1.
\label{biorth}
\end{equation}
Then the divergent part of ${\bf W}^{-1}$ in a neighborhood of $E_{\nu}$ is 
nothing but the truncation,
\begin{equation}
{\bf W}^{-1}_{tr}=\frac{\Lambda_{\nu}\, \tilde \Lambda_{\nu}'}
{\tilde \Lambda_{\nu}'\, {\bf W}(E)\, \Lambda_{\nu}}\, ,
\label{trunc}
\end{equation}
where there is an explicit dependence on $E$ in the denominator. This 
denominator, a number, vanishes at $E=E_{\nu}.$ As a matrix element of
${\bf W}$ it is nothing but the Wronskian of the following two waves,
$F \equiv {\bf f}^+\, \Lambda_{\nu}'$
and
$\xi \equiv {\Phi}\, \Lambda_{\nu}.$
The former, $F,$ is irregular, the latter, $\xi,$ is regular. While 
$\Lambda_{\nu}$ and $\Lambda'_{\nu}$ do not depend on $E,$ since they were 
defined at $E=E_{\nu},$ both $F$ and $\xi$ depend on $E,$ via ${\bf f}^+$ and
${\bf \Phi}.$ When their Wronskian vanishes, $F$ and $\xi$ become proportional 
to each other, and there exits a number $c$ such that $F_{\nu}=c\, \xi_{\nu}.$ 
This special wave is both a mixture of regular solutions and a mixture of Jost 
solutions, with positive imaginary parts in the momenta driving all Jost 
solutions. Therefore it decreases exponentially in all channels when 
$r \rightarrow \infty$ and it is square integrable as well as regular. As 
expected it represents either a bound state or a regularized resonance.

\medskip
According to Eqs.(\ref{residue1},\ref{trunc}), the residue under study comes 
from just the reciprocal of the derivative of the Wronskian of $F$ and $\xi,$
\begin{equation}
{\cal R}_{\nu} = \frac{\Lambda_{\nu}\, \tilde \Lambda_{\nu}'}
{  
d \left[\tilde \Lambda_{\nu}'\, {\bf W}(E)\, \Lambda_{\nu}\right]/{dE}
 \,  |_{E=E_{\nu}} 
}\, .
\label{residue2}
\end{equation}
In short, we must calculate the derivative of a Wronskian with respect to 
the energy, $d \left[\tilde \Lambda_{\nu}'\, {\bf W}(E)\, \Lambda_{\nu}\right]
/dE.$ To help manipulations with Wronskians, 
define an operator matrix ${\bf U}$ with matrix elements the CSM potentials, 
completed by the centrifugal barriers and the thresholds,
\begin{equation}
{\bf U}_{ij}=e^{2i\theta}U_{ij}\left( e^{i\theta} r \right) + \delta_{ij}
\left[e^{2i\theta}E_j^*+\frac{\ell_j(\ell_j+1)}{r^2} \right].
\end{equation}
Then elementary, but slightly tedious manipulations, which are already 
described in \cite{Newt} or in Appendix B of \cite{us}, 
give the remarquably simple result,
\begin{equation}
d \left[\tilde \Lambda_{\nu}'\, {\bf W}(E)\, \Lambda_{\nu}\right] /
dE\, |_{E=E_{\nu}} = -\, c \,
\int_0^\infty dr\, \tilde \xi(E_{\nu},r)\, \xi(E_{\nu},r).
\label{Euclid}
\end{equation}
Then the constant $c$ cancels out between this and the numerators of 
${\bf f}^+_\nu\,\tilde {\cal R}_{\nu}\,\tilde {\bf \Phi}_{\nu}$ 
and
${\bf \Phi}_{\nu}\,{\cal R}_{\nu}\,\tilde {\bf f}^+_\nu,$
which make the same, symmetric formula anyway, whether $r > r'$ or $r < r',$ 
since $F_{\nu}=c\, \xi_{\nu}.$ 

\medskip
Summing upon all such residues obtained at roots $E_{\nu}$ of 
$\det {\bf W}$ above the ``opener'' curves in the $P$ upper half-plane, the 
contour integral reads,
\begin{equation}
{\cal I}(r,r') = -\, 2\, i \, \pi\, \sum_{\nu} \ 
\frac{                        {\bf \Phi}(E_{\nu},r)\ \Lambda_{\nu}\ 
\tilde \Lambda_{\nu}\ \tilde {\bf \Phi}(E_{\nu},r')          } 
{\int_0^{\infty}dr''\, 
\tilde \Lambda_{\nu}\ \tilde {\bf \Phi}(E_{\nu},r'')\ 
                             {\bf \Phi}(E_{\nu},r'')\ \Lambda_{\nu} } \, .
\label{residue3}
\end{equation}
Here we state again that the column vector $\Lambda_{\nu}$ is the null, 
right-hand side eigenvector of ${\bf W}(E_{\nu}),$ namely 
${\bf W}(E_{\nu})\, \Lambda_{\nu}=0,$ then the 
column vector ${\bf \Phi}(E_{\nu})\, \Lambda_{\nu}$  of wave functions is the 
wave function of the bound state or resonance, and the denominator plays the 
role of a ``Euclidean-like square norm''. This denominator is non vanishing; 
this corresponds to the hypothesis of single, isolated poles. All these are 
labeled by $\nu,$ a discrete index, or as well by $P_{\nu},$ 
an isolated root of ${\bf W}$ if viewed as a function of $P.$

\subsection{Completeness}

\medskip
Since the three contributions ${\cal I}_{sc},$ $\sum_j{\cal I}_j$ and 
${\cal I}$ are obviously related by ${\cal I}_{sc} + \sum_j{\cal I}_j = 
{\cal I},$ it is trivial to equate $\frac{i}{2\pi}{\cal I}_{sc},$ the 
multichannel identity, with the difference between  $\frac{i}{2\pi}{\cal I},$ 
the pseudoprojector on both bound states and resonances, and  $\frac{i}{2\pi}
\sum_j {\cal I}_j,$ the latter term making the pseudoprojector upon the 
continuum for all channels. Naturally, in practical calculations, a cutoff and 
some amount of discretization will be necessary to integrate such continuum 
terms, but the $P$ representation provides a suitable frame for testing the 
convergence of such a resolution for sum rules, level densities and similar
observables. Notice that, because of the use of complex, self dual bras 
and kets in the resolution, such cutoff and discretization manipulations
may generate spurious imaginary parts for the expectation values of hermitian
observables. For a discussion and possible interpretation of imaginary 
parts in individual matrix elements, we refer to \cite{Berg2}. But, when 
summed upon all discrete and integral terms provided by the resolution,
such imaginary parts must add up to a negligible, spurious noise compared to
the real parts. This requested cancellation makes one more criterion to 
validate numerical operations.

\section{Discussion And Conclusion}

\medskip
Once again we used the ABC theorems \cite{ABC} to locate the discrete 
spectrum at trivially rotated positions deduced from the discrete spectrum
of an initial, hermitian Hamiltonian. The topological similitude provided
by the CSM rotation warrants that, as long as there are no double poles or
higher singularities with the initial Hamiltonian, the same will be true with
the CSM Hamiltonian.

\medskip
Then it was not very difficult to find a representation which allows a suitable
contour integration of the Green's function. There was still a slightly 
complicated Riemann surface to handle, for the number of cuts was reduced to 
$N-2$ only \cite{Weid}, but we took great care, including a few numerical, 
illustrative examples, to show that all cuts in the new representation are 
well understood, all thresholds are easily located, all complex momenta to be 
used for proofs have positive imaginary parts in a physical domain of a 
suitable sheet, and in general that all technicalities are sound.

\medskip
This proof of the CSM completeness for $N$ channels is restricted to a finite 
number of well separated channels, normal square root threshold singularities,
in a purely inelastic situation, without rearrangement, and with short ranged 
forces. The case of long range forces makes a more difficult question, indeed 
\cite{atomists1} \cite{atomists2}. But our restrictions still allow a large 
class of practical problems, and for instance in nuclear physics, a very large 
number of collective resonances can be described by the coupled channel 
equations that we studied.

\bigskip \noindent
Acknowledgment: B.G.G. thanks the Hokkaido University for its hospitality 
during part of this work.

\newpage
\centerline{Appendix}

\bigskip
We give here in some detail a description of that regular solution $\phi_{.j}$
which accounts for the discontinuity of the Green's function across a cut. 
For the sake of pedagogy, we set the channel number to be $N=4$ and shall
consider only what happens for, e.g., the second cut. Generalizations are 
obvious and left as an exercise for the interested reader. In a condensed 
notation, we write the upper rim Wronskian matrix as,
\begin{equation}
{\bf W}_u=\left[\matrix{a & b & c & d \cr 
                  e & f & g & h \cr 
                  i & j & k & l \cr 
                  m & n & o & p }\right].
\end{equation}
where, for instance, $b$ is the Wronskian of $f^+_{.1}$ with $\varphi_{.2}$ and
$o$ is the Wronskian of $f^+_{.4}$ with $\varphi_{.3}.$ The inverse of 
${\bf W}_u$ reads, trivially,
\begin{equation}
{\bf W}_u^{-1}=(det_u)^{-1}\left[\matrix{a' & e' & i' & m' \cr 
                                   b' & f' & j' & n' \cr 
                                   c' & g' & k' & o' \cr 
                                   d' & h' & l' & p' }\right],
\label{cofact}
\end{equation}
where $det_u$ is the determinant of ${\bf W}_u$ and the prime symbols denote 
the corresponding cofactors. For the lower rim of the second cut a 
substitution occurs for the second row of ${\bf W}_u,$ hence the lower rim 
Wronskian matrix reads,
\begin{equation}
{\bf W}_l=\left[\matrix{a & b & c & d \cr 
                  q & r & s & t \cr 
                  i & j & k & l \cr 
                  m & n & o & p }\right],
\end{equation} 
where, for instance, $t$ is the Wronskian of $f_{.2}^-$ with $\varphi_{.4}.$
Accordingly the inverse matrix becomes,
\begin{equation}
{\bf W}_l^{-1}=(det_l)^{-1}\left[\matrix{a'' & e' & i'' & m'' \cr 
                                   b'' & f' & j'' & n'' \cr 
                                   c'' & g' & k'' & o'' \cr 
                                   d'' & h' & l'' & p'' }\right],
\end{equation}
where doubleprime symbols denote new cofactors, but the cofactors of 
$\{q,r,s,t\}$ are the same as those of $\{e,f,g,h\}.$

\medskip
Again with a transparent, condensed notation, we set, for the upper and lower 
rim, respectively,
\begin{equation}
\tilde {\bf f}^u=\left[\matrix{A & B & C & D \cr 
                               E & F & G & H \cr 
                               I & J & K & L \cr 
                               M & N & O & P }\right], \ \ \ \
\tilde {\bf f}^l=\left[\matrix{A & B & C & D \cr 
                               Q & R & S & T \cr 
                               I & J & K & L \cr 
                               M & N & O & P }\right].
\end{equation}
with, for instance, $\{A,B,C,D\} \equiv 
\{f^+_{11},f^+_{21},f^+_{31},f^+_{41}\},$ and $\{E,F,G,H\} \equiv 
\{f^+_{12},f^+_{22},f^+_{32},f^+_{42}\},$ while
$\{Q,R,S,T\} \equiv \{f^-_{12},f^-_{22},f^-_{32},f^-_{42}\}.$ For $r < r',$
the discontinuity to be studied corresponds to the transposed of 
Eq.(\ref{discont}), and reads, in a condensed notation,
\begin{equation}
\tilde {\bf D}_2(r')={\bf W}_u^{-1}\,  \tilde {\bf f}^u(r')-
                     {\bf W}_l^{-1}\,  \tilde {\bf f}^l(r'),
\label{discontt}
\end{equation}
The subscript $2$ for the cut and the $r'$ dependence will be now 
understood and we shall use trivial identities to analyze
\begin{equation}
\tilde {\bf D}=\left[{\bf W}_l^{-1} +
{\bf W}_l^{-1}\, (\Delta {\bf W})\, {\bf W}_u^{-1}\right]\, \tilde {\bf f}^u -
{\bf W}_l^{-1}\,  (\tilde {\bf f}^u + \Delta \tilde {\bf f}) = 
{\bf W}_l^{-1}\, (\Delta {\bf W})\, {\bf W}_u^{-1}\, \tilde {\bf f}^u - 
{\bf W}_l^{-1}\, \Delta \tilde {\bf f},
\end{equation}
where $\Delta {\bf W} = {\bf W}_l - {\bf W}_u$ and 
      $\Delta \tilde {\bf f} = \tilde {\bf f}^l - \tilde {\bf f}^u.$ The 
point is, both modifications $\Delta$ are just substitutions for second rows; 
they boil down to dyadics,
\begin{equation}
\Delta {\bf W} = \left[\matrix{0 \cr 1 \cr 0 \cr 0} \right]\  \otimes\ 
\left[ q-e\ \ \ \ r-f\ \ \ \ s-g\ \ \ \  t-h \right],\ \ \ \ \ \ 
\Delta \tilde {\bf f} = \left[\matrix{0 \cr 1 \cr 0 \cr 0} \right]\  \otimes\ 
\left[ Q-E\ \ \ \ R-F\ \ \ \ S-G\ \ \ \  T-H \right].
\end{equation}
(Our use of the tensor product symbol $\otimes$ is actually superfluous; we 
just want to stress the matrix product of a column by a row.)
The next point is, then, that a global dyadic form for $\tilde {\bf D}$ 
emerges,
\begin{equation}
\tilde {\bf D}={\bf W}_l^{-1}\, \left[\matrix{0 \cr 1 \cr 0 \cr 0} \right]
\otimes \left(\ \left[q-e\ \ \ \ r-f\ \ \ \ s-g\ \ \ \ t-h\right]\, 
{\bf W}_u^{-1}\, \tilde {\bf f}^u - 
\left[Q-E\ \ \ \ R-F\ \ \ \ S-G\ \ \ \ T-H\right]\, \right).
\end{equation}
Furthermore, from the very definition of matrix inversion, we see that
\begin{equation}
\left[e\ \ \ \ f\ \ \ \ g\ \ \ \ h\right]\, {\bf W}_u^{-1}=
\left[0\ \ \ 1\ \ \ 0\ \ \ 0\right],
\end{equation}
hence
\begin{equation}
\left[-e\ \ \ \ -f\ \ \ \ -g\ \ \ \ -h\right]\, {\bf W}_u^{-1}\, 
\tilde {\bf f}^u = - \left[E\ \ \ F\ \ \ G\ \ \ H\right],
\end{equation}
and $\tilde {\bf D}$ simplifies into
\begin{equation}
\tilde {\bf D}=(det_l)^{-1}\, \left[\matrix{e' \cr f' \cr g' \cr h'} \right]
\otimes \left(\ \left[q\ \ \ \ r\ \ \ \ s\ \ \ \ t\right]\, {\bf W}_u^{-1}\, 
\tilde {\bf f}^u - \left[Q\ \ \ \ R\ \ \ \ S\ \ \ \ T\right]\ \right).
\end{equation}
For $r < r'$ the complete discontinuity ${\bf \Phi}(r)\, \tilde {\bf D}(r')$ 
of ${\bf G}(r,r')$ thus reads
\begin{equation} 
det_l\ det_u\ {\bf \Phi}(r)\ \tilde {\bf D}(r')=\phi(r)\, \tilde \Xi(r'),
\end{equation}
with
\begin{equation}
\phi=e'\, \varphi_{.1}+f'\, \varphi_{.2}+g'\, \varphi_{.3}+h'\, \varphi_{.4},
\end{equation}
and
\begin{equation}
\Xi=(qa'+rb'+sc'+td')\, f^+_{.1}+(qe'+rf'+sg'+th')\, f^+_{.2} +\ 
   ...\  +(qm'+rn'+so'+tp')\, f^+_{.4}- det_u\, f^-_{.2}.
\label{tricky0}
\end{equation}
Both $\phi$ and $\Xi$ are column vectors and relate to the second cut, hence
they should actually read $\phi_{.2}$ and $\Xi_{.2}$ in a notation 
compatible with Eqs.(\ref{continuum},\ref{resola}). We omitted such 
subscripts, for the sake of conciseness.

\medskip
It may be convenient to take advantage of the cofactor nature of all the
coefficients $a',$ ... $p'.$ This gives indeed the formal, but condensed 
formula,
\begin{equation}
\phi=\det\, \left[ \matrix{ 
      a      &       b       &       c      &      d       \cr 
\varphi_{.1} & \varphi_{.2}  & \varphi_{.3} & \varphi_{.4} \cr
      i      &       j       &       k      &      l       \cr 
      m      &       n       &       o      &      p       } \right].
\end{equation}
Similarly, we find the formal result,
\begin{equation}
\Xi= - \det\, \left[ \matrix{ 
      a      &       b       &       c      &     d    &    f^+_{.1}  \cr 
      e      &       f       &       g      &     h    &    f^+_{.2}  \cr
      i      &       j       &       k      &     l    &    f^+_{.3}  \cr 
      m      &       n       &       o      &     p    &    f^+_{.4}  \cr
      q      &       r       &       s      &     t    &    f^-_{.2}} \right],
\label{tricky1}
\end{equation}
because all coefficients such as,
\begin{equation}
qa'+rb'+sc'+td'=\det\, \left[ \matrix{ 
      q      &       r       &       s      &     t      \cr
      e      &       f       &       g      &     h      \cr
      i      &       j       &       k      &     l      \cr 
      m      &       n       &       o      &     p     } \right],
\ \ ...\ \ , \ \  
qm'+rn'+so'+tp'=\det\, \left[ \matrix{ 
      a      &       b       &       c      &     d      \cr
      e      &       f       &       g      &     h      \cr
      i      &       j       &       k      &     l      \cr 
      q      &       r       &       s      &     t     } \right],
\label{tricky2}
\end{equation}
can themselves be interpreted, after keeping track of signs, as cofactors for 
the last column of the determinant shown by Eq.(\ref{tricky1}).

\medskip
In the $2N$-dimensional space of solutions, it is known that the $2N$ Jost 
solutions and the $N$ regular ones are related by a formula such as,
\begin{equation}
{\bf f}^-={\bf \Phi}\, {\bf W}^{-1}\, {\bf w} + 
{\bf f}^+\, {\bf w}^{-1}\, {\bf W}_-\, {\bf W}^{-1}\, {\bf w},
\label{convert}
\end{equation}
where ${\bf W}$ is the same as ${\bf W}_u,$ while ${\bf W}_-$ is the analog
of ${\bf W}$ if one replaces each $f_{.m}^+$ by its partner $f_{.m}^-$\, .
Then ${\bf w}$ is a diagonal matrix, defined from the Wronskians 
${\cal W}\left(f_{.m}^+,f_{.n}^-\right)=-2ik_m\delta_{mn}.$ It will be 
noticed from Eq.(\ref{convert}) that, if we expand an $f^-_{.m}$ on the 
basis spanned by all the $\varphi_{.n}$ and all the $f^+_{.n}$\, , the regular 
components of $f^-_{.m}$ are provided by the $m$-th column of the matrix 
product ${\bf W}^{-1}\, {\bf w}.$

\medskip
It is known that $\Xi$ always belongs to the subspace of $N$ regular 
solutions. In our illustrative example where $N=4$ and we studied the second 
cut, our $\Xi,$ according to Eq.(\ref{tricky0}), is a superposition of five 
solutions, namely all the $f_{.n}^+$ and one $f^-_{.n}$ only, $ f^-_{.2}.$ 
After an expansion of $f_{.2}^-$ on the basis spanned by the $\varphi_{.n}$ 
and the $f^+_{.n},$ all its irregular components must cancel out those 
preexisting irregular components of $\Xi$ seen from Eq.(\ref{tricky0}). (For 
the sake of rigor, we verified, by brute force calculations when $N=2,$ $3$ 
and $4,$ that the components $f^+_{.n}$ do vanish out.)
Thus we may consider the regular components only, coming from just $f_{.2}^-.$ 

\medskip
The weight of $f_{.2}^-$ is, according to Eq.(\ref{tricky0}), $-det_u.$ 
We must therefore find the second column of,
\begin{equation}
-det_u\, {\bf W}^{-1}\, {\bf w}=-\left[\matrix{a' & e' & i' & m' \cr 
                                               b' & f' & j' & n' \cr 
                                               c' & g' & k' & o' \cr 
                                               d' & h' & l' & p' }\right]\,
                                 \left[\matrix{-2ik_1 & 0 & 0 & 0 \cr 
                                               0 & -2ik_2 & 0 & 0 \cr 
                                               0 & 0 & -2ik_3 & 0 \cr 
                                               0 & 0 & 0 & -2ik_4 }\right],
\end{equation}
hence the final result,
\begin{equation}
\Xi=2ik_2\, (e'\, \varphi_{.1} + f'\, \varphi_{.2} + g'\, \varphi_{.3} + 
h'\varphi_{.4}) = 2ik_2\, \phi.
\end{equation}
The generalization, $\Xi_{.j}=2ik_j\, \phi_{.j}$ is obvious. For any channel
number $N$ and any $j$-th cut, both $\phi$ and $\Xi$ correspond to
the $j$-th column of ${\bf W}^{-1},$ hence to the cofactors of the $j$-th row
of ${\bf W}.$ There is no need here to specify ${\bf W}_u$ or ${\bf W}_l,$
because the relevant cofactors are the same on both rims of the cut. The fact 
that $\Xi$ and $\phi$ are the same except for the factor $2ik_j$ gives
the same, symmetrical result whether $r$ is larger or smaller than $r'.$ 
And the denominator present in Eqs.(\ref{continuum},\ref{resola}) reads, when 
all factors are collected,
\begin{equation}
{\cal D}(E)=\frac{det_u\, det_l}{2ik_j}\, .
\end{equation}


\begin{thebibliography}{99}

\bibitem{ABC} 
Aguilar, J. and Combes, J.M. {\it Commun. Math. Phys.} {\bf 22} (1971) 269;
Aguilar, J. and Balslev, E. {\it Commun. Math. Phys.} {\bf 22} (1971) 280

\bibitem{Kato1} 
Myo, T. Ohnishi, A. and Kat\=o, K. {\it Progr. Th. Phys.} {\bf 99} (1998) 801

\bibitem{Kato2} 
Myo, T., Kat\=o, K., Aoyama, S. and Ikeda, K. {\it Phys. Rev. C} {\bf 63} 
(2001) 054313-1

\bibitem{Berg1} 
Berggren, T. {\it Phys. Lett.} {\bf B44} (1973) 23;
Romo, W. J. {\it Nucl. Phys.} {\bf A237} (1975) 275

\bibitem{us}
Giraud, B.G. and Kat\o, K. {\it Ann. Phys.} {\bf 308} (2003) 115

\bibitem{KFFGR} 
Krylov, N.S. and Fock, V.A. {\it Zh. Eksp. Teor. Fiz.} {\bf 17} (1947) 93;
Fonda, L., Ghirardi, G.C. and Rimini, A. {\it Rep. Prog. Phys.} {\bf 41} 
(1978) 587

\bibitem{ML}
Menon, V.J. and Lagu, A.V. {\it Phys. Rev. Lett.} {\bf 51} (1983) 1407

\bibitem{BRT}
Barrett, R.F., Robson, B.A. and Tobocman, W. {\it Rev. Mod. Phys.} {\bf 55} 
(1983) 155

\bibitem{R} Robson, D. pp. 179-248 in {\it Nuclear Spectroscopy and 
Reactions}, J.Cerny ed., Academic Press N.Y. and London (1975)

\bibitem{Newt} Newton, R.G. {\it Scattering Theory of Waves and Particles} 
(1982) Springer Verlag (New-York, Heidelberg, Berlin) ISBN 0-387-10950-1,
3-540-10950-1; see in particular chapters 12, 15 and 17.

\bibitem{Berg2} Berggren, T. {\it Phys. Lett.} {\bf B373} (1996) 1

\bibitem{Weid} Weidenm\"uller, H.A., {\it Ann. of Phys.} {\bf 28} (1964) 60

\bibitem{atomists1}
Ho, Y.K. {\it Phys. Reports} {\bf 99} (1983) 1;  see also the reference list
of this paper

\bibitem{atomists2} 
Moiseyev, N. {\it Phys. Reports} {\bf 302} (1998) 211; see also the 
reference lists of this paper

\end{thebibliography}
\end{document}